\documentclass{article}
\usepackage[table]{xcolor}
\usepackage[numbers]{natbib}
\usepackage{graphicx}
\usepackage{amssymb}
\usepackage{todonotes}
\usepackage{booktabs}
\usepackage{multirow}
\usepackage[affil-it]{authblk}
\usepackage[hyphens]{url}
\providecommand{\keywords}[1]{\textbf{\textit{Index terms:}} #1}
\usepackage{listings}

\usepackage{color}

\definecolor{codegreen}{rgb}{0,0.6,0}
\definecolor{codegray}{rgb}{0.5,0.5,0.5}
\definecolor{codepurple}{rgb}{0.58,0,0.82}
\definecolor{backcolour}{rgb}{0.95,0.95,0.92}

\lstdefinestyle{mystyle}{
    backgroundcolor=\color{backcolour},
    commentstyle=\color{codegreen},
    keywordstyle=\color{magenta},
    numberstyle=\tiny\color{codegray},
    stringstyle=\color{codepurple},
    basicstyle=\footnotesize,
    breakatwhitespace=false,
    breaklines=true,
    captionpos=b,
    keepspaces=true,
    numbers=left,
    numbersep=5pt,
    showspaces=false,
    showstringspaces=false,
    showtabs=false,
    tabsize=2
}

\begin{document}
\title{Hydras and IPFS: A Decentralised Playground for Malware}




\author{Constantinos Patsakis and Fran Casino}
\affil{University of Piraeus, Greece}


\date{}
\maketitle

\begin{abstract}
Modern malware can take various forms, and has reached a very high level of sophistication in terms of its penetration, persistence, communication and hiding capabilities. The use of cryptography, and of covert communication channels over public and widely used protocols and services, is becoming a norm.
In this work, we start by introducing Resource Identifier Generation Algorithms. These are an extension of a well-known mechanism called Domain Generation Algorithms (DGA), which are frequently employed by cybercriminals for bot management and communication. Our extension allows, beyond DNS, the use of other protocols. More concretely, we showcase the exploitation of the InterPlanetary file system (IPFS). This is a solution for the ``\textit{permanent web}'', which enjoys a steadily growing community interest and adoption. The IPFS is, in addition, one of the most prominent solutions for blockchain storage. We go beyond the straightforward case of using the IPFS for hosting malicious content, and explore ways in which a botmaster could employ it, to manage her bots, validating our findings experimentally. Finally, we discuss the advantages of our approach for malware authors, its efficacy and highlight its extensibility for other distributed storage services.
\end{abstract}


\section{Introduction}
The continuous arms race \cite{MANSFIELDDEVINE201815} between malware authors and security practitioners has led to the existence of a wide range of tools and techniques which have turned cybercrime into a profitable ``business''. This business has various monetisation sources, ranging from spamming \cite{rao2012economics} to ad injection \cite{CHEN2017164} but also including, notably, denial of service, ransomware-based extorsion and phishing. Monetisation schemes were in the past challenging and frequently problematic for the attacker, as traditionally the adversary needs to (at least partly) disclose her identity to receive the money. Yet, the recent availability of private cryptocurrencies (such as Monero, Zcash, etc.) has significantly decreased this exposure.

However, a second risk of deanonymization still exists for the cybercriminal, as she needs to maintain control over the infected devices or bots. It is clear that a classical scheme where the adversary (botmaster) issues commands and distributes them in a centralised fashion has inherent single-point-of failure issues that impede its success. For instance, if there is a unique domain or IP that is used to control the infected devices, once it is identified it will be (quickly) blocked. Therefore, botmasters try to either hide their commands within legitimate network traffic that cannot be blocked, e.g. social networks or widely used services, or to decentralise botnet management. It should be highlighted that a typical botnet consists of several thousands of devices which, on top of traditional workstations may consist of routers, printers, smartphones and other smart devices.

\begin{figure*}[!ht]
    \centering
    \includegraphics[width=\textwidth]{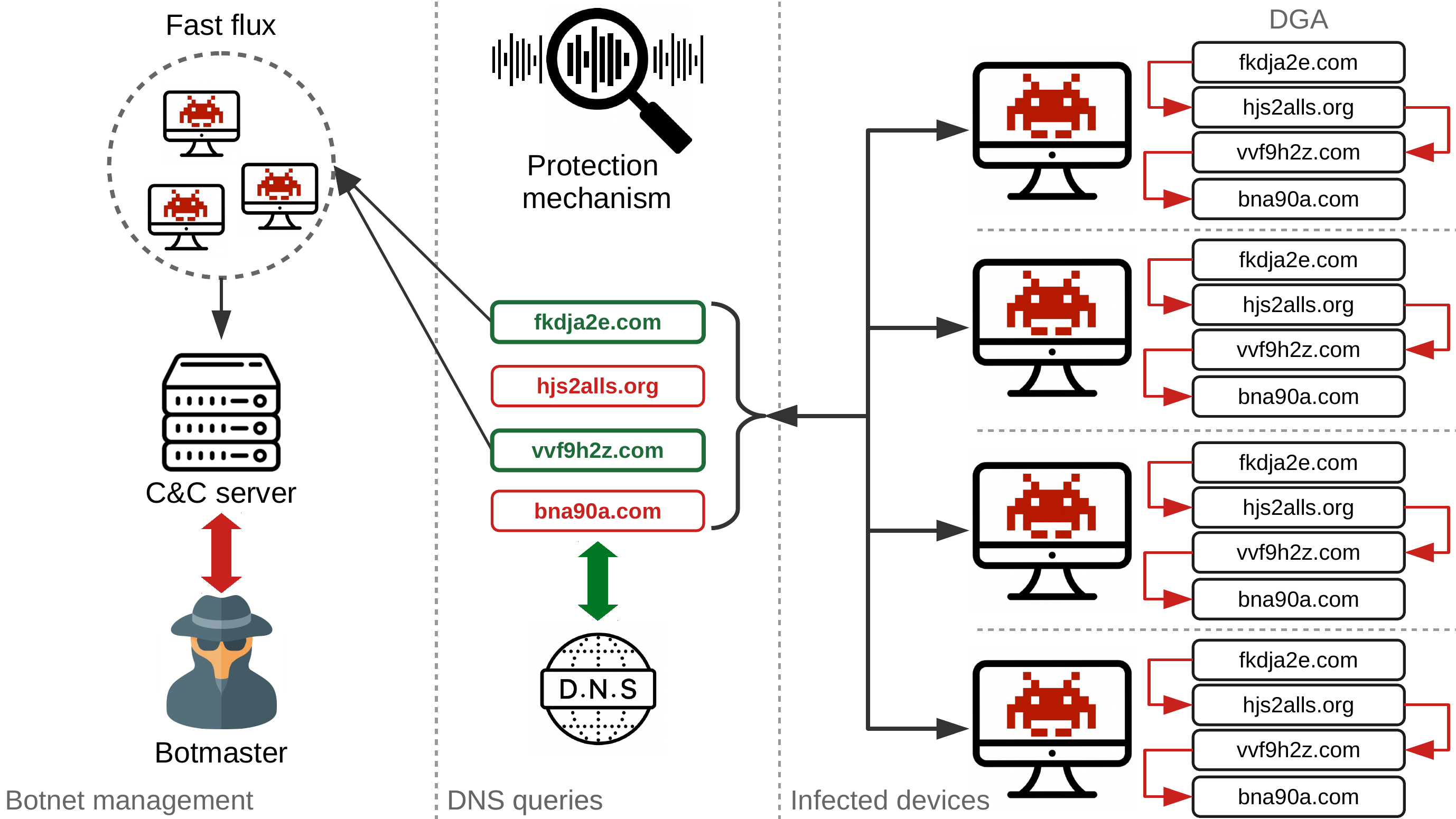}
    \caption{The typical DGA management model.}
    \label{fig:dga_model}
\end{figure*}

A typical strategy that cybercriminals use to counter both deanonymization and blacklisting issues is the use of multiple domains to manage infected hosts. In this context, bots run Domain Generation Algorithms (DGAs) which generate multiple domains at a predefined time (generally each day) which will all be queried by the bots to retrieve commands. In this regard, the attacker may generate thousands of bogus domains but limit her control to only a handful of them. However, this means that the control is periodically (e.g. daily) transferred from one domain to another without any specific pattern for the defendant's perspective and with each generated domain having an equal probability of being used by the attacker. The problem is amplified by the lack of proper reporting from domain registrars, who often fail to timely respond to requests from law enforcement authorities trying to take down botnets or at least trace the identity of malware authors. In specific cases, registrars have been even found to misbehave, e.g. by accepting bribes \cite{krebs2010mariposa}. To further hide their identity and avoid blacklisting, botmasters may also loop through a set of infected hosts to host the content that the DGAs point to. Therefore, the problem resembles the mythical Lernaean Hydra which spawned two heads when one of her heads was chopped. The typical management model used by botmasters when using DGAs is illustrated in Figure \ref{fig:dga_model}.

\subsection{Main contributions}

Pondering on the new directions that malware can have, the goal of this work is to investigate the use of the InterPlanetary File System (IPFS)\footnote{\url{https://ipfs.io}} for the coordination of a botnet. While we argue that the use of IPFS for the distribution of malicious content is relatively clear, we discuss other issues that emerge from this perspective. More precisely, we argue that the fact that IPFS can provide anonymity, persistence of the content, fast delivery, and a robust network where content cannot be easily blocked, provides an ideal landscape for malware authors. It should be noted that IPFS has already started being used maliciously. Quite recently, IPFS was used during a phishing campaign for hosting phishing forms\footnote{\url{https://www.bleepingcomputer.com/news/security/phishing-attacks-distributed-through-cloudflares-ipfs-gateway/}}. In this regard, we investigate IPFS-based botnets, their applicability and issues that emerge from their possible use.

To the best of our knowledge, this is the first work that provides a holistic architecture to discuss the exploitation of IPFS from botnets, beyond the trivial scenario of just storing malicious or illegal content. More concretely, we investigate an IPFS address generation algorithm which uses a skewed PRNG to generate hashes that will eventually point to files that contain a new set of instructions to be executed by the infected devices. Apparently, in the proposed setting we do not have new domain names but new IPFS addresses, however, due to the way IPFS works it is the name resolution. The use of IPFS implies a series of inherent properties and benefits for the adversary compared to similar works which rely on the use of social networks or blockchain \cite{nagaraja2011stegobot,moubarak2018developing,princeflashback,Ali2018,8433199}. Our proposed methodology is further distributed, more robust and provides more bandwidth. Contrary to traditional DGAs, the IPFS flavour that we discuss in this work is more stealth since it hides the traces of the botmaster and provides a better distribution network with no apparent take-down mechanism.

Based on the above, we consider that DGAs and the methodology used for IPFS can be considered different instantiations of a broader concept. Therefore, we introduce the concept of a \textit{Resource Identifier Generation Algorithm (RIGA)}. We argue that the growing trend towards decentralised systems, the introduction of new protocols with different content resolution methods will eventually be exploited from malware to disseminate the content. Since an algorithm could generate the names that can be used from different protocols the notion of RIGA encapsulates the core concept of DGAs: generate more output than the actual resource cannot be determined from an outsider. Therefore, a RIGA is a natural extension of DGAs that considers other protocols.

\subsection{Ethical considerations} The goal of this work is not to inspire malware authors to write more efficient malware but to motivate researchers to find solutions for an emerging threat. Unarguably, due to its nature IPFS can be considered ideal for spreading malware and illegal/malicious content since among others there is little regulation, a lot of anonymity, and the content can be considered permanent. The fact that many big organisations like Cloudflare are backing the IPFS initiative means that IPFS is here to stay. To this end, there is a need to study how an adversary could exploit this new technology and prepare possible countermeasures and initiate a dialogue with the IPFS community to see how can similar issues be resolved in the near future. The research in this field must, therefore, investigate aspects beyond the trivial ones,, e.g. hosting malicious content or forms, which as already discussed have been already observed. To this end, the proof of concept code only provides the simple implementation of a DGA that is not hardened to provide further security to the adversary.

\subsection{Organisation of this work}
The rest of this work is structured as follows. In the next section we provide an overview of the related work focusing mainly on DGAs and IPFS. In Section \ref{sec:riga}, we extend the notion of DGAs and introduce the concept of RIGAs. Then, in Section \ref{sec:proposed} we detail the proposed scheme for decentralised management of bots through IPFS. Section \ref{sec:experiments} illustrates the efficacy of our proposal through various experiments. In Section \ref{sec:discussion}, we compare our work to the current state of the art. Finally, we conclude this work discussing future directions and open issues.

\section{Related work}
\label{sec:related}
Nowadays, most malware-based campaigns rely on the use of botnets, commanded by remote servers (i.e. C\&C servers), which send instructions/orders to infected devices \cite{203628}. In the past, the mechanism used to find such C\&C servers was to hardcode IP addresses in the malware. However, such technique entailed a set of drawbacks for the attackers \cite{nadji2017still} (e.g. easy to take-down). Therefore, botnets evolved into peer-to-peer (P2P) botnets \cite{grizzard2007peer}, which adopted a myriad of techniques such as the \textit{Fast Flux} approach, which imitates content distribution networks by resolving a domain name to multiple IP addresses \cite{holz2008measuring,akamai}.

On top of the previously discussed approaches, malware use DGAs, which implement a deterministic pseudo-random generator (PRNG) to create a set of domain names \cite{7535098,6175908}. Hence, infected devices check the list of generated domains and perform queries until they find the C\&C server, whose location may also change dynamically. In this scenario, blacklisting domains is rendered useless as it implies many practical issues.

\subsection{Domain Generation Algorithms}

According to  Plohmann et al. \cite{197187}, DGAs are categorised as:
\begin{itemize}
    \item Arithmetic-based DGAs
    \item Hash-based DGAs
    \item Wordlist-based DGAs
    \item Permutation-based DGAs
\end{itemize}

In general, arithmetic-based DGAs use a PRNG to combine a set of characters (typically ASCII) and create a domain name. In the case of hash-based DGAs, attackers basically create domains using the hex representation of a hash. To detect both DGA-based families, methods reported in the literature use features such as entropy, length or lexical characteristics to determine whether a DGA has generated a domain name or not \cite{Aviv2011,6151233}, as well as characteristics such as traffic information (e.g. NXDomain queries or WHOIS information) \cite{Zhou2013DGABasedBD,5762763,1,yadav2012,yadavgraph,gongodyseey,7163279}.

Wordlist-based DGAs appeared to overcome the drawbacks of the previous two DGA families. In this case, attackers adopt the use of English wordlists to generate ``human-readable'' domains, hindering previous DGA detection approaches. In this context, several neural network-based techniques exhibit good detection accuracy \cite{18,Anderson2016}, as well as other novel methods based on metrics such as n-grams and word feature extraction  \cite{curtin2018detecting,stefanotracking}.

In the case of permutation-based DGAs, attackers can use combinations based on an initial domain to bypass detectors \cite{johannesbader}. There exist similar techniques such as \textit{domain shadowing}, which relies on the use of valid domains that were previously hacked \cite{Liu2017}.

Regardless of the DGA used, the typical botnets use Internet protocols for communication. Nevertheless, more sophisticated an original approaches can be found in the case of social networks \cite{nagaraja2011stegobot,princeflashback} and blockchain \cite{Ali2018,8433199,moubarak2018developing}, which entail further detection challenges, since all traffic seems legitimate and is covered under standard channels such as HTTPS. In the case of blockchain networks, the problem is exacerbated due to its inherent properties such as immutability and anonymity.

For a detailed overview and classification of methods of how malicious domains can be detected, the interested reader may
refer to \cite{zhauniarovich2018survey}.


\subsection{IPFS}
\label{sec:ipfs}

The InterPlanetary File System (IPFS) \cite{benet2014ipfs} is a distributed P2P system for retrieving and sharing IPFS objects. IPFS uses a Merkle Directed Acyclic Graph (DAG), which is a cryptographically authenticated data structure, to address such objects. Therefore, instead of identifying objects by their location (e.g. HTTPS), the system addresses them by their representation of the content itself, which is usually their Base58 SHA-256 encoded hash\footnote{\url{https://en.bitcoin.it/wiki/Base58Check_encoding#Base58_symbol_chart}}.

The contents of an IPFS object are mainly stored in two fields: (i) the data field, which is an unstructured binary data block of size of 256 KB (i.e. larger files are stored by a list of links to file chunks that are $<$ 256 kB) and (ii), an array of links to other IPFS objects (e.g. other files under the same directory), which are used to increase the network efficiency. An example of the contents of an IPFS object is depicted in Listing \ref{lst:codeIPFS}. Note that the two first characters of the hash (i.e. Qm) encode the hash algorithm used and its length according to the multihash\footnote{\url{https://multiformats.io/multihash/}} format.

\begin{lstlisting}[language=Python, label=lst:codeIPFS, caption=Overview of an IPFS object structure.]
> ipfs object get QmYWAifyw2V5... | jq
{
  "Links": [...some links...],
  "Data": "binary data blocks"
}
\end{lstlisting}


Similar decentralised systems such as BitTorrent can successfully coordinate the transfer of data between millions of nodes, but it applies only to the torrent ecosystem. In contrast, IPFS implements a generalised version of this protocol called \texttt{BitSwap}, which enables further possibilities such as built-in storage marketplaces like Filecoin\footnote{\url{https://filecoin.io/filecoin.pdf}}.

Filecoin is a distributed electronic currency that uses proof-of-retrievability, which is a verification mechanism used to prove that a node stores a particular file. In this context, currency is awarded for storing files, which is a practice that is gaining more adepts (e.g. Ethereum's Swarm and Mist or MaidSafe and the SAFE network)\footnote{\url{https://www.ibtimes.co.uk/juan-benet-ipfs-talks-about-filecoin-1586122}}.

The use of a Merkle DAG structure allows the creation of a version control system (VCS). More concretely, IPFS stores the object history so that all versions are accessible throughout time. This permits configurable synchronisation of files since all users can edit them locally and later push the new files to IPFS. VCS is enhanced by the InterPlanetary Naming System (IPNS), which enables content linking, so that files can be accessed using the node ID address, allowing users to retrieve updated contents without knowing the new hashes of such files. Another relevant characteristic of IPFS is that it is a self-certifying file system (SFS), which means that data served to clients is authenticated by their own filename and the node providing it. Therefore, nodes use their private keys to ``sign'' data objects they publish, and the authenticity of this data can be verified using their public-key. In addition to these features, IPFS enables a series of properties described briefly in Table \ref{tab:properties}.

 \begin{table}[h]
   \rowcolors{2}{gray!25}{white}
   \setlength{\tabcolsep}{9pt}
   \scriptsize
   \caption{IPFS main properties.}
  \begin{tabular}{p{.2\columnwidth}p{.65\columnwidth}}
     \toprule
   \textbf{Property} &  \textbf{Description}  \\
   \midrule
Immutability/ Tamper proof       &  Content-based addressing guarantees that each file resolves to a specific hash\\
  Equality  & All peers have similar, if not equal, permissions and possibilities. \\
      Decentralisation       &  The network is totally distributed with no central entities \\
      Fault tolerance        & A high number of individual peers guarantees the robustness of the network   \\
      Availability  & The availability of the network depends on multiple peers and not on a single entity\\
  Resilient to attacks  &  The large number of peers guarantee the persistence of the network \\
  Unlimited Resources & A high number of simultaneous users sharing their assets. \\
Scalability & Requests are made to the closest peer instead to a single central location, avoiding bottlenecks \\
 Marketplace monetisation       &  With systems like Filecoin, which incentivise IPFS \\
     \bottomrule
   \end{tabular}
   \label{tab:properties}
 \end{table}

Nowadays, the applications and possibilities of IPFS are being exploited in a myriad of contexts \cite{HUCKLE2016461,kellywayback16,benet2014ipfs}. Of particular relevance is the symbiotic relationship between IPFS and blockchain, a distributed and linked immutable ledger \cite{nakamoto2008bitcoin}. Therefore, the combination of both technologies enhances their application scenarios \cite{swan2015blockchain,Casino2018AIssues} by enabling off-chain storage and anonymous file sharing, which are usually managed through smart contracts \cite{szabo1997idea}.

\section{Resource Identifier Generation Algorithms}
\label{sec:riga}
The core concept of a DGA is to create a pseudo-random sequence of domains that the botmaster may use to host his malicious content. Due to the vast amount of domains that can be generated, the defendant cannot blacklist all the domains, nor can she determine which one will be used by the botmaster. While in DGAs this concept is implemented through the DNS protocol, one could extend this to accommodate for other protocol and approaches.

To this end, we extend the notion of a DGA to that of a \textit{Resource Identifier Generation Algorithm (RIGA)}. In principle, a RIGA generates a sequence of possible addresses that can be used by an adversary to host malicious content. We assume that we have an address resolution space where the adversary can host her desired malicious content, and each address has the same probability of hosting it for an external observer.

The sequence of these addresses can be purely random or pseudo-random, depending on the size of the address resolution space. For instance, if the address resolution space is minimal, then the adversary knows that with the proper query rate the RIGA would eventually generate the desired addresses. Clearly, in this approach, the RIGA does not need any input. However, in the case of a large address resolution space, the adversary needs to know that the RIGA will produce a set of predefined addresses which will be generated by all infected devices. Therefore, the RIGA takes as input a parameter which will be used as a seed by a PRNG.

\begin{figure*}
    \centering
    \includegraphics[width=\textwidth]{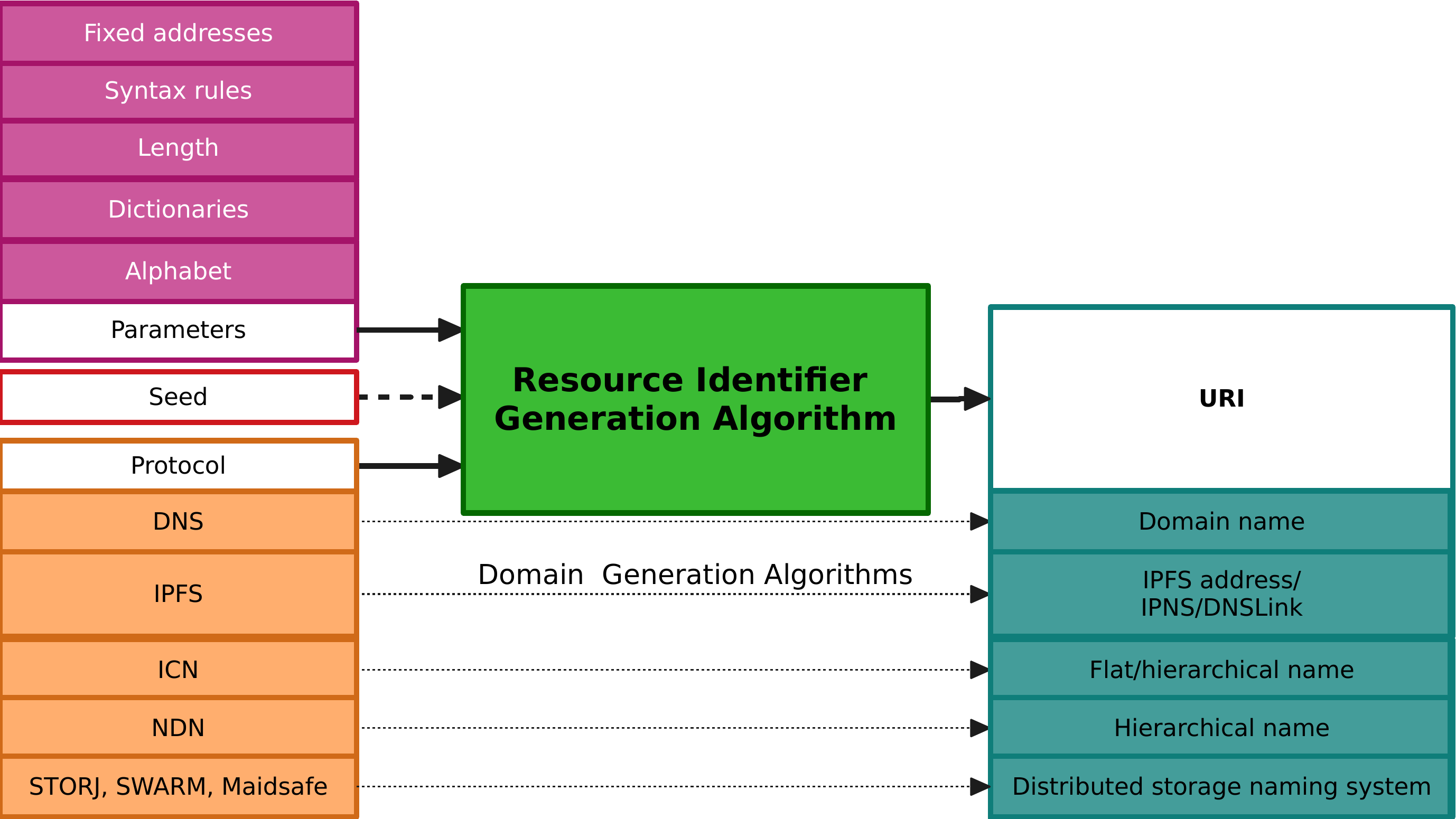}
    \caption{An overview of the potential scope of Resource Identifier Generation Algorithms.}
    \label{fig:riga}
\end{figure*}

Regardless of whether the RIGA will use a PRNG or true random numbers, the RIGA has a mapping mechanism to convert a number into a Uniform Resource Identifier (URI) of the corresponding protocol, which will be its output. In the case of the DNS protocol, the RIGA generates a series of URLs some of which the adversary will allocate to host her content. In the case of IPFS, the RIGA could generate a series of IPFS, IPNS or DNSLink addresses, as it will be discussed in the next section.
Similarly, RIGAs could be used to generate URIs for SWARM \cite{swarm}, STORJ \cite{storjvsipfsbruno}, Maidsafe \cite{maidsafe} and other distributed storage services. Finally, a RIGA could generate hierarchical or flat names for Information-Centric Networks, or hierarchical names for the case of Network Defined Networks.  An overview of the potential scope of RIGAs and its relationship with DGAs is illustrated in Figure \ref{fig:riga}.

While the protocol may imply some parameters, a RIGA would need some additional modifications to generate the desired output. Typical parameters needed for a RIGA include the length of the desired output and the alphabet that will be used. For instance, the URL encoding implies a specific alphabet which differs from the one used in IPFS. Moreover, one RIGA may use some dictionaries to produce the names or part of them. For instance, the TLDs in all DGAs are hard-coded in their source code, while some RIGAs may use dictionaries to produce the names,, e.g. the Matsnu DGA. Some RIGAs may require specific syntax rules on the alphabet to create a meaning name,, e.g. Base58Check encoding or use the letters of the alphabet with specific frequencies of occurrence,, e.g. the Symmi DGA. Finally, one may require the RIGA to generate a set of predefined URIs without having to hard-code them in the source code. More details about the latter will be discussed in the next section.

\section{Proposed scheme for IPFS}
\label{sec:proposed}
In the next paragraphs, we describe the way that IPFS can be exploited by malware focusing on how DGAs can use it. Therefore, the infection mechanism to spread the malware that penetrated the systems is not discussed and is considered beyond the scope of this work. The infection mechanism depends on the vulnerabilities that these devices might have, the fact that they are web discoverable and accessible with default credentials or that the user has been tricked to visit malicious URL. Note that the same method could be extended for other distributed storage services like STORJ, SWARM etc. as outlined in the previous section. For the sake of clarity, in what follows we will only focus on IPFS.

First, we provide a brief overview of the proposed botnet management model with IPFS-based DGAs, then we detail the new DGA algorithm and provide a skewed PRNG for its construction. Afterwards, we discuss seeding and spreading issues and the effectiveness of takedown mechanisms. Finally, we close this section providing a discussion for upstream communication.

\subsection{Overview}
Given the continuous growth of malware, in what follows we discuss the possibility of exploiting IPFS. The reason behind this choice is that we have seen several transformations of the mondus operandi of modern malware, see Section \ref{sec:related}. Therefore, the growing community of IPFS and the seamless procedure of uploading and sharing content is for sure expected to attract people willing to exploit the provided features for nefarious means. In fact, these considerations have already sparked several discussions in the IPFS community on, e.g. how to treat illegal content and whether deletion of content should be made possible\footnote{See \url{https://github.com/ipfs/faq/issues/9}
\url{https://github.com/ipfs/faq/issues/36}
\url{https://github.com/ipfs/faq/issues/156}}. While many opt for the blacklisting approach to counter this issue, we argue that this may only partially address these issues. The reason is more or less similar to the one discussed about the regular web, if the addresses that have to be blacklisted are far too many, then this approach is rendered useless since it implies a lot of continuous effort from the participating nodes. To this end, we discuss the possibility of having IPFS-based RIGAs.

The use of IPFS to armour a botnet enables a set of opportunities for malware authors including immutability, removal of costs, easiness to spread, and anonymity. In general, using IPFS, the bots can disseminate the commands of the C\&C server only by pinning a file. Moreover, once infected, they can create different versions of the instruction file and spread a new ``version'' of the malware, which translates into a resilient and mutable botnet. By default, the traffic in IPFS is encrypted, and HTTPS can be used to prevent protocol censoring in, e.g. corporate environments. Even more, since nobody manages the contents nor the traffic, implementing filters or detection schemes is more difficult than in the case of centralised services, and it is impossible to know all the available files unless their hash is known.

Clearly, no one can know beforehand whether a hash link will be used or not. Even more, having access to the original source code of the RIGA is useless as the content may become almost immediately available. Therefore, one cannot determine whether any of the generated hashes will be uploaded. Moreover, since the hash is not reversible, one cannot determine the content of the shared files. Contrary to DNS, the botmaster does not have any cost, and there is no direct link with her identity. The adversary, from now on Malory, can upload her files whenever she wants and they will be immediately distributed without an obvious take-down mechanism. 

\subsection{Modus operandi}
Based on the above, Malory prepares some files with the malicious context, or links to that context in the form of files,, e.g. rendezvous points. The latter approach allows her to set the content at a later stage. Having these files, she will upload them to IPFS when she deems appropriate. To allow her bots to access these files she uses a skewed PRNG; which will be detailed afterwards, that she embeds in the malware of her campaign. Using a predefined seed, the PRNG will iterate through random IPFS links, which most likely will not point to any content. However, the PRNG will also generate the predefined links at the desired input.

To allow the bots retrieve the content that Malory has prepared when accessing content beyond IPFS,, e.g. the rendezvous points, Malory embeds further information in the malware. This information includes the instruction set that bots can execute and how it can be retrieved. For instance,  the malicious content can be embedded using steganographic methods in images in social networks to avoid raising suspicions. Therefore, the bots are given the instructions on (i) how to collect the information from IPFS, (ii) how to parse retrieved content, (iii) how to navigate to a set of rendezvous points, (iv) how to extract the content from these points, (v) how to extract the commands from the collected content, and (vi) how to provide her feedback in the rendezvous points. Clearly, the content that Malory distributes could be digitally signed, so Malory would have to embed a set of public keys. This would allow her bots to verify that Malory submitted the command and avoid take over/down efforts.

\begin{figure*}[!ht]
    \centering
    \includegraphics[width=\textwidth]{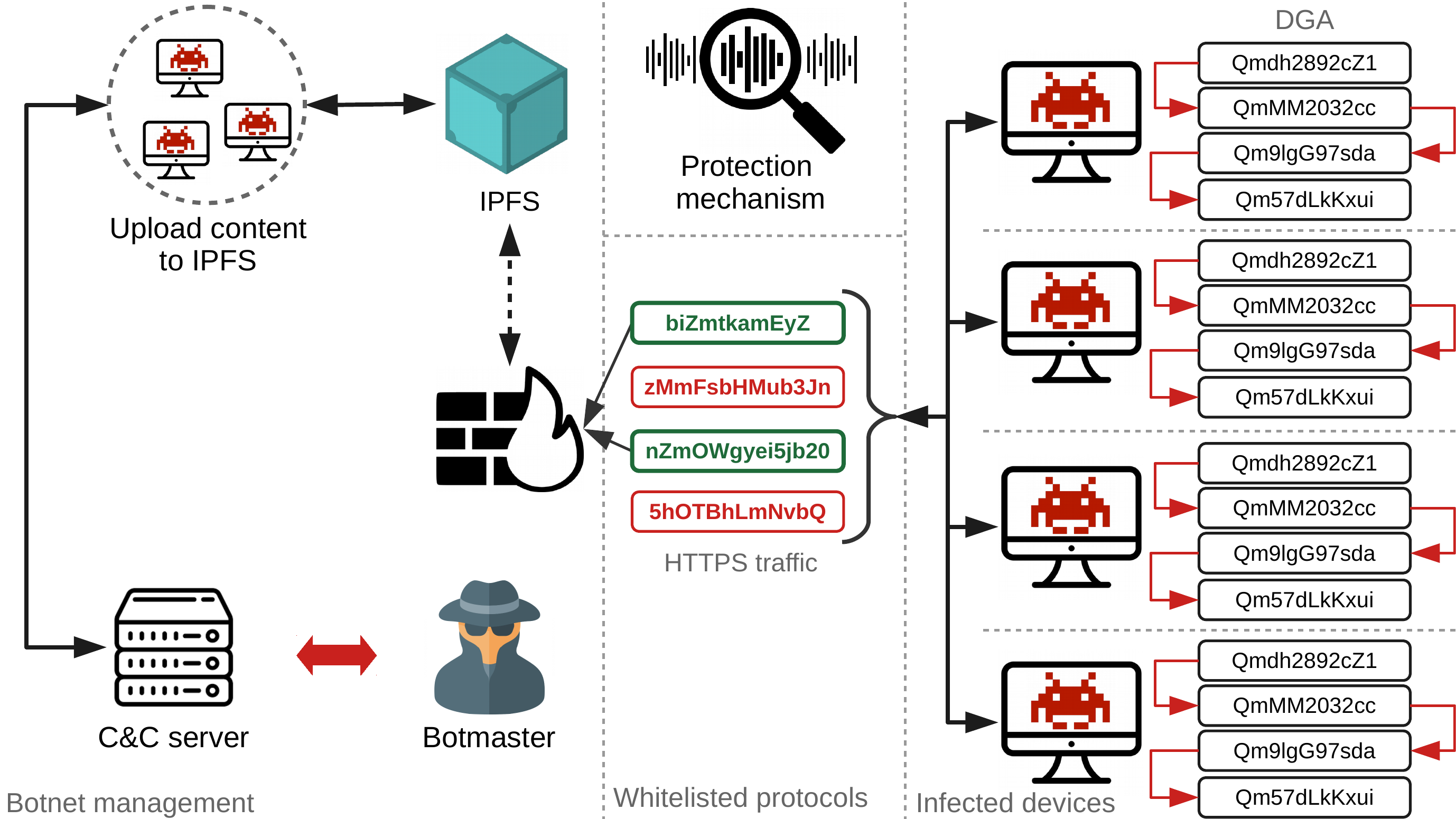}
    \caption{Proposed management model.}
    \label{fig:proposed}
\end{figure*}

From the bots' side, Malory has instructed them to query IPFS files periodically using the RIGA. To this end, Malory has embedded the list of well-known IPFS gateways\footnote{\url{https://ipfs.github.io/public-gateway-checker/}} which the bots use according to a round-robin schedule. The latter is used to minimise the risk of alerts and blacklisting in the event of having short intervals between the queries. Bots use the RIGA, generate a domain and try to access it via one of the available domains using a predefined timeout to guarantee that the search for content will not result into an endless loop if the content does not exist (see Figure \ref{lst:code}). If the content exists, the bot retrieves it and executes the listed commands or redirects itself to a new location to retrieve them according to the retrieval algorithm that Malory has embedded. The concept behind redirection is that Malory may not have already decided her actions so she will upload them at a later stage, but she cannot change the IPFS hash. Therefore, she prepares a placeholder to do so at a later stage. The proposed management model is illustrated in Figure \ref{fig:proposed}.

\begin{lstlisting}[language=Python, label=lst:code, caption=Example code to request content from IPFS with a timeout.]
import requests
h = "Qmc8N5wtMkvMySqxu4Agy2SGv"
h = h + "L2zxYGf4rWmHvMASoUQv6"
r = requests.get('https://ipfs.io/ipfs/%s'%h, timeout=5)
\end{lstlisting}

\subsection{A PRNG for an IPFS RIGA}
\label{sec:hashcreation}
In a typical DGA, the author creates a pseudo-random number generator (PRNG) that generates the domains and selects a small subset of the possible values and registers the corresponding domains. However, in the IPFS scenario, this is not possible as IPFS uses cryptographic hash functions and the result is encoded to Base58 encoding. Therefore, it is computationally impossible to select some hashes and create files that hash into these values.

To achieve the same effect, we create a skewed PRNG which generates a set of predefined hashes that will be used as pointers for links to content which be used by bots to resolve new Malory's commands. In what follows we represent this PRNG as a function $f$ which iterates through various states. For the sake of simplicity, we assume that we have a counter $x\in\{0,1,2,...,U\}=\mathbb{D}$ that the bots progressively increase starting from $0$. While we consider $0$ as the initial seed, this can differ as the seed may be collected dynamically.

Moreover, we assume that Malory has prepared a set of $k$ files and/or links to content (also stored as files). Malory selects a hash function $h(x):\mathbb{D}\rightarrow [0,2^{L}-1]$,, e.g. SHA-256, and hashes the files with it creating $h_i,i\in\{1,2,...,k\}$. Then, Malory selects the values of the counters when the PRNG will evaluate to her desired values that we denote as $v_i,i\in\{1,2,...,k\}$. Therefore, Malory requires from the PRNG to evaluate as follows:
\[
f(v_i)=h_i,\forall i\in\{1,2,...,k\}
\]
Since the scope of this work is to discuss the possible use of IPFS-based RIGAs, we deliberately omit the hardening of the PRNG. In this context, one of the most obvious ways to compute $f$ is using polynomial interpolation. Hence, we compute the polynomial in terms of Lagrange polynomials as follows:
\[
    f(x)=\sum_{i=0}^n\left(\prod_{0\leq j\leq n, j\neq i}\frac{x-x_j}{x_i-x_j}\right)h_i\bmod p
\]
where $p$ is a prime number with $p>h_i,\forall i\in\{1,2,...,k\}$.

Clearly, evaluating $f$ over $v_i,i\in\{1,2,...,k\}$ will produce the desired values $h_i,i\in\{1,2,...,k\}$, and the rest of the values will be random values in the range of $[0,2^L-1]$. Since IPFS represents the hash of files and objects using Multihash format and Base58 encoding, the DGA will iterate through the values of $\mathbb{D}$ and evaluate the polynomial at a predefined rate and convert the values from integers in the interval $[0,2^L-1]$ to the corresponding Base58 encoding which can be directly associated with an IPFS link.

It is evident that the methodology above can be used for any hash-based naming scheme like IPFS. A trivial implementation of this PRNG is available on GitHub\footnote{\url{https://github.com/kpatsakis/RIGA}}.


While the RIGA above efficiently generates the domains, it does not exploit all the features of IPFS. More precisely, since all objects in IPFS are content-addressed every mutation of an object leads to a new address, IPFS provides two protocols to create mutable addresses in this ecosystem, IPNS and DNSLink with the latter being more efficient. Thus, by having the IPNS or DNSLink address of an object one may constantly access the newest version of an object.

The fact that the protocols above allow one to access mutable content can be exploited by Malory in another way. In this scenario, each bot may parse $\mathbb{D}$ randomly and evaluate a function $f'$ similar to $f$ which now generates IPNS links. Malory can update her content whenever she wants, and the bots will access it in a random sequence. The use of IPNS enables further possibilities, such as linking malicious files (e.g. during a brief window of time) and then substituting them by new non-malicious content, so that no traces are left. Moreover, old content can be unpinned and hence removed by the participating nodes creating further issues in the analysis and gaps in the timeline of the campaign.

\subsection{Content seeding and spreading}

In IPFS, nodes and files can be uniquely identified. In the case of nodes, their ID corresponds to the hash of their public key. In the case of files, identification is performed by computing the hash of their content, as stated in Section \ref{sec:ipfs}. By default, each node is connected to a swarm of nodes which can be retrieved with the command \texttt{ipfs swarm}. However, it is not possible to retrieve the list of all existing nodes in IPFS nor the files stored by each node. Therefore, files are only accessible by users that know their corresponding hashes. Despite these limitations, the list of nodes that store a file can be retrieved using the command \texttt{ipfs dht findprovs <hash>}. Thus, nodes can become seeders of a file by pinning it using \texttt{ipfs pin add <hash>}.

The anonymity of nodes and files can be seen as a challenge to malware spreading. Note that when uploading content to IPFS, the node that uploads it becomes its primary and unique seeder. If a malicious node acts as the unique C\&C server, it can be easily detected and isolated. To counter this deficiency, a subset of infected devices may install IPFS and automatically pin the shared file from Malory, immediately becoming seeders of this content. Practically, this means that the botnet has a robust infrastructure which cannot be taken down due to its decentralised nature. As already discussed, currently there is no mechanism to delete a file in IPFS and no plan to support this functionality in the near future.

It should be noted that Malory can improve the seeding of her content and obscure the identity of the infected nodes that participate in her campaign by motivating other nodes to share her content using Filecoins. Since the content that Malory would request to share would be minimal in terms of size, at the proper cost, many nodes could be easily convinced to share the proposed content. As a result, Malory has the guarantees not only from the subset of infected hosts but from seemingly benign hosts that her content will be disseminated.







\subsection{Take down mechanisms}
In the traditional DGA scenario, organisations have often joined their efforts to take down botnets. One of the most well-known examples is Conficker for which international cooperation was made that managed to registered all the domains that the DGA would generate, preventing the botmaster from contacting the bots and eventually lose their control.

Evidently, this is not possible in the IPFS scenario since all the links are actually hashes of files from cryptographic hash functions. Therefore, even if the malware has a ``\textit{kill switch}'' (as in the case of Wannacry) that can be activated by a remote command that it receives through an IPFS link, no one can do this for an IPFS link. The latter would require one to generate a file with a specific hash that would be generated from the DGA. Since cryptographic hash functions are immune to pre-image attacks, this approach is impossible.

Similarly, traffic to IPFS might not be easily blocked. The fact that several gateways can be used means that one must restrict access to all IPFS facing content. Given the momentum, while this option sounds practical, it is not rational due to the increasing content that is shared. It should be highlighted that with continuous use of IPFS in conjunction with blockchains, blocking IPFS content in several corporate environments becomes inapplicable.

\subsection{Receiving bots' feedback}
While the infected devices have to retrieve commands from the botmaster, bots must also return some information to her. This information may be trivial,, e.g. bot $ID_X$ has entered the network, or may contain valuable information,, e.g. credit cards, credentials etc. While this information can easily be stored in IPFS, it is impossible for the botmaster to retrieve it as she cannot guess the hash of the information to retrieve it.

Therefore, the botmaster may use IPFS to publish rendezvous points that the bots will use to push information back to the botmaster by either posting it directly in the rendezvous point or the IPFS link where they have uploaded it. In the latter case, infected nodes may later unpin such files, leaving no trace of what information was exfiltrated. Note that such files will be accessed only by the botmaster who will receive the hash from the rendezvous point. Obviously, as in the case of Malory's commands, this information can be encrypted or embedded using steganography in another object before sharing.

\section{Experimental results}
\label{sec:experiments}

To validate the feasibility of our approach, we implemented two tests. The first one tries to determine whether the gateways implement any measure to block frequent host requests. The scope of this experiment is to see whether a host that will have an IPFS-based RIGA would be blocked after often requests to a gateway. The second experiment tries to quantify how much time it would take Malory to make her content available to her bots.

Once we checked the efficiency of the network by locally pinning a set of files (i.e. content is available in the order of seconds, so it enables real-time) our first test focused on studying whether the gateways apply any threshold on client requests. Therefore, we selected 20 Wikipedia articles which are hosted on IPFS and collected their addresses. Then, we collected the list of available IPFS gateways. For each gateway, we sequentially tried to fetch each article 50 times from the same host. Note that we added a maximum timeout for each query to be resolved since IPFS does not implement an efficient timeout mechanism in the case of non-existing files. The results for each gateway are illustrated in Table \ref{tbl:ipfs_wiki}. In the reported results, dropped refers to requests that timed out. From the table, it is evident that the gateways do not respond uniformly. Therefore, there are significant differences in the amount of time needed for each gateway to respond and there are discrepancies even with the different timeouts of the requests. Apart from the \url{swedneck.xyz} gateway which had by far the longest time to respond to the requests and with relatively high timeouts (3.1\% and 5.4\%) all the other gateways almost never timed out (4 timeouts in total). Practically, the above results indicate that IPFS-based RIGAs can perform requests to IPFS gateways at a rate of one request per 2 seconds with a timeout of 3 seconds facing no throttling issues from the gateways. Clearly, in the case of native IPFS protocol, there are no such considerations; however, we examine the HTTPS approach of the infected devices.

\begin{table*}[!ht]
    \centering
    \footnotesize
    \begin{tabular}{l|rr|rr}
        \toprule
        \multirow{2}{*}{        \textbf{Gateway}}& \multicolumn{2}{c}{\textbf{5 sec timeout}} & \multicolumn{2}{c}{\textbf{3 sec timeout}}\\
         & \textbf{Time} & \textbf{Dropped}& \textbf{Time} & \textbf{Dropped}\\
        \midrule
        \url{https://ipfs.io/ipfs/} &373.492 & 0& 422.889 &0 \\
         \url{https://gateway.ipfs.io/ipfs/}& 391.009 & 0  & 374.111  &0 \\
         \url{https://ipfs.infura.io/ipfs/}& 854.953 & 0 & 1123.762 & 0 \\
         \url{https://xmine128.tk/ipfs/}& 381.595 & 0 & 383.206 & 0\\
         \url{https://ipfs.jes.xxx/ipfs/}& 438.0490 &0 & 1903.150 & 1\\
         \url{https://siderus.io/ipfs/}& 280.261& 0 &296.138 & 0\\
         \url{https://www.eternum.io/ipfs/}& 594.157&  1 & 609.238  & 0\\
         \url{https://hardbin.com/ipfs/} & 457.364  &0& 601.464 & 0 \\
         \url{https://ipfs.wa.hle.rs/ipfs/} & 1234.263 &0 & 1043.500  &0\\
         \url{https://ipfs.renehsz.com/ipfs/} & 1482.931& 0  & 482.282 & 0 \\
\url{https://cloudflare-ipfs.com/ipfs/} &285.893  &1 & 289.622  & 1\\
    \url{https://ipns.co/} & 1848.695 & 0  & 1143.398 &0 \\
    \url{https://gateway.swedneck.xyz/ipfs/} &  5952.236  &31 & 5887.395 & 54\\
         \bottomrule
    \end{tabular}
    \caption{Statistics for 1000 requests of Wikipedia articles using different IPFS gateways. Time (in milliseconds) refers to the total time for all requests.}
    \label{tbl:ipfs_wiki}
\end{table*}

\begin{figure}
    \centering
    \includegraphics[width=.5\textwidth]{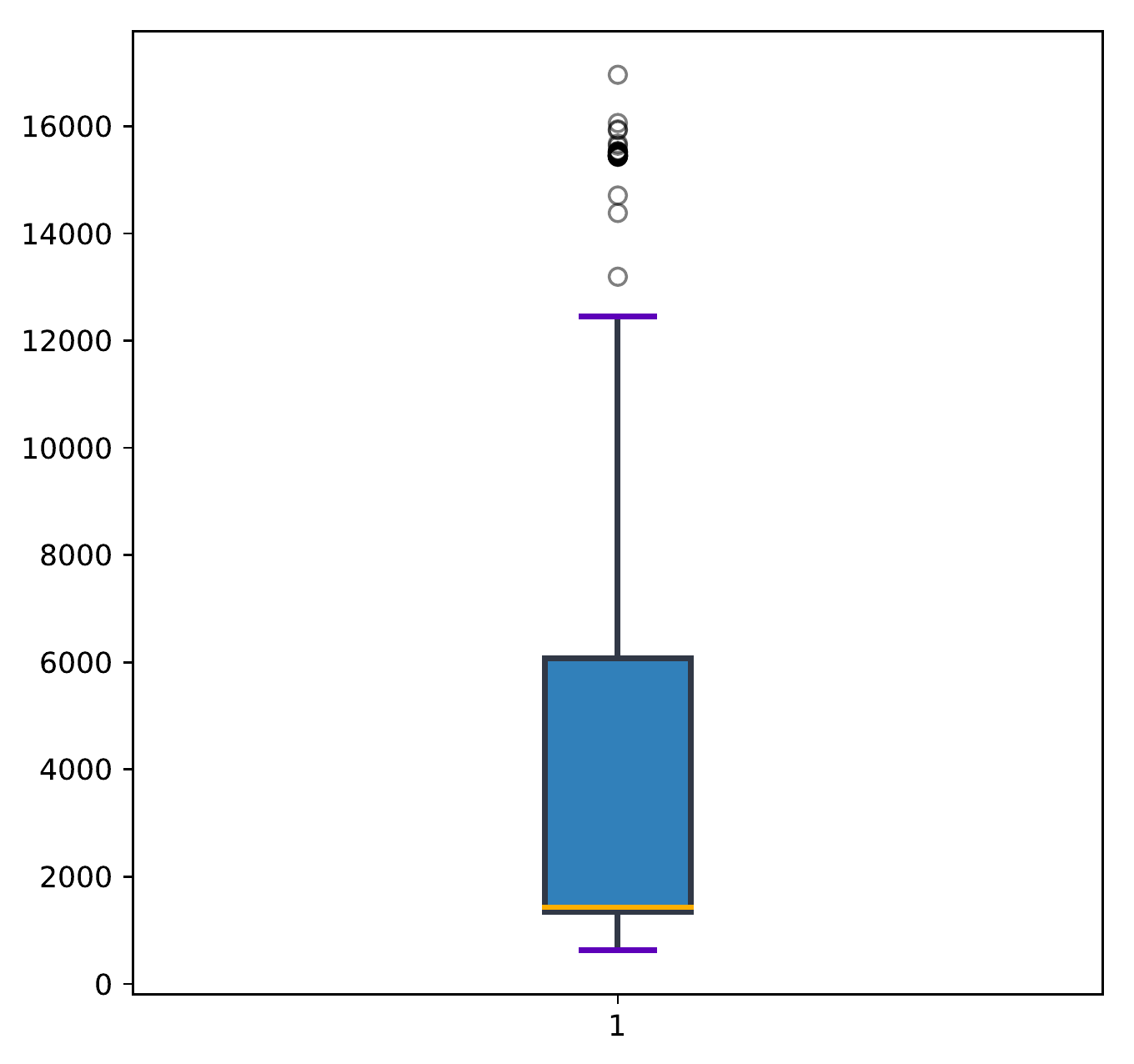}
    \caption{Boxplot of the time results (in milliseconds) for estimating the time required for a file to become available on IPFS.}
    \label{fig:boxplot}
\end{figure}

As already discussed, the second experiment aims to determine the time required for content to become available in IPFS. To this end, we upload a total of 1000 files to IPFS with a 4KB size and, later, we retrieve them using the four most efficient gateways, according to Table \ref{tbl:ipfs_wiki} (i.e. \url{https://cloudflare-ipfs.com/ipfs/}, \url{https://xmine128.tk/ipfs/}, \url{https://siderus.io/ipfs/} and \url{https://gateway.ipfs.io/ipfs/}). In each case, we use a relaxed timeout of 5 seconds to get the files, randomly shifting to another if one fails to timely deliver it. Therefore, we capture the timestamp when adding the files to IPFS and after retrieving them and compute their difference. Note that the aforementioned procedure bypasses any possible locality bias since the file is actually requested from a remote IPFS gateway. The average time and standard deviation of the full procedure were $3647ms$ and $3715ms$, respectively. As illustrated in Figure \ref{fig:boxplot}, there are several outliers which significantly increase the average value and standard deviation. Nevertheless, the results indicate that the availability is well bounded in the scale of just seconds, enabling real-time adaptable malware campaigns, where malicious content can be rapidly spread.


\section{Discussion}
\label{sec:discussion}
Compared with the current state of the art, the proposed methodology of RIGAs over IPFS has several advantages.
As in many approaches, the proposed methodology is totally decentralised, providing robustness, availability, and high scalability. Moreover, contrary to many approaches, our IPFS methodology establishes connections with whitelisted domains over encrypted channels, using a standard and whitelisted protocol (HTTPS), and with legitimate and trusted certificates. A typical example is Cloudflare, one of the biggest content delivery networks which acts as an exit node for IPFS. Note that due to the continuous adoption of blockchain even the direct use of IPFS protocol is expected to be whitelisted in a wide range of devices quite soon, as it is one of the most viable solutions for storing data on the blockchain. It is worth noting that Aquino et al. explored the potential exploitation of IPFS for distributing the commands of a C\&C server in \cite{wscdc}. However, they only consider the static scenario where the bot finds the content uploaded on IPFS as it would in the case of other approaches e.g. using social networks.

As already discussed, a botnet with the proposed methodology is not easy to take down. This can be attributed to the actual nature of IPFS. Despite its decentralised architecture, IPFS not only does not have a deletion mechanism, but its community is not eager to support it. The blacklisting approach which has been proposed in the community to address malicious content is not certain that every node in the network will use that. Moreover, when the size of the blacklisted content grows beyond a point, the additional overhead is expected to deter many nodes from using it. Therefore, given the proper motivation (e.g. Filecoins) or by making infected devices to pin content, the shared content is expected to stay on the network and not be taken down.

Since IPFS node IDs are created using the hash of their public key (generated using a 2048-bit RSA) a node ID can be reset multiple times. Moreover, IPFS objects do not store which nodes seed them, and thus, it is not possible to relate a file with its initial seeder, which hinders tracking mechanisms. In this regard, IPNS remote pinning services such as Textile\footnote{https://www.textile.io/} implement useful mechanisms which enable decentralised content management with high availability, enhancing the privacy and anonymity of peers. Nevertheless, although developers are working to enhance network's anonymity, there is still a big gap to cover. For example, since the architecture of IPFS connects nodes by their proximity to enhance network efficiency, using Tor connections would disclose the user's private, public and .onion addresses information. Notwithstanding, content retrieving from a gateway using Tor connections is anonymous \footnote{https://dweb-primer.ipfs.io/avenues-for-access/lessons/tor-gateways.html}. Other features such as end-to-end integrity can also be satisfied utilising of ipfs-sec domains and DNSSEC so that users do not need to trust intermediate gateways \footnote{\url{https://blog.cloudflare.com/e2e-integrity/}}, a property analogous to that provided by HTTPS (i.e. with Internet Service Providers (ISP) acting as intermediate gateways).

The use of blockchains, as proposed in by Ali et al. \cite{Ali2018} may also provide a decentralised approach on  permanent storage; however, their proposed solution allows the communication is small chunks, and therefore, many messages have to be broadcast from the botmaster. Taking into consideration the low rate of verified commits of the Bitcoin network, the bots will have to wait far longer than the reported values. It should also be noted that bots would have to continually monitor the blockchain which implies a significant processing and unnecessary overhead. On the contrary, the IPFS approach can deliver content of arbitrary size almost instantly and without having to monitor the network for updates continually.

In the IPFS implementation of RIGAs, contrary to DGAs, the adversary does not have an additional cost that may disclose her identity. More precisely, in DGAs the adversary must purchase some domain names from registrars. In this regard, the adversary may perform the transactions using stolen cards and fake identities, however, the cost can be substantial if the amount of domains is large. Nevertheless, the IPFS approach does not imply any cost for the adversary and no substantial wait period. For instance, when purchasing a domain name, depending on the registrar, there might be an idle period until the domain name becomes available which may range from hours to days. In the IPFS approach though there is no idle period and content become almost immediately available.

Finally, it should be noted that the IPFS approach is more stealth. In traditional DGAs, the requested domain name due to its entropy or WHOIS data may disclose that it is an output of a DGA. However, in the IPFS case, this is not relevant. The requests are made to legitimate and encrypted domains. Therefore, this will not leave any traces in the DNS queries.

\section{Conclusions}
\label{sec:conclusions}
Cybercrime has become a very profitable ``business'' having various monetisation sources. Modern  malware creators bypass most of current security architectures by using, among others, cryptographic methods  and covert communication channels. In this work, we explore the possibilities of extending DGAs and then focus on the context of a recent decentralised data storage system, namely IPFS.

To this end, we first extend the notion of DGAs into a more generic framework, namely RIGA, which is a family of mechanisms to generate a sequence of possible addresses that can be used to host malicious content, regardless of the context. Next, we showcase the exploitation of the IPFS network to enable bot management and malware spreading through the corresponding instantiation of a RIGA. Our experimental results show that IPFS could be effectively exploited to convey such an attack, due to the lack of defence mechanisms in terms of query resolution (gateways), the lack of an effective deletion mechanism, and its speed, enabling real-time campaigns. Moreover, we provide a functional PRNG implementation of an IPFS-enabled RIGA. Finally, we discuss the advantages of our approach for malware authors to raise awareness and motivate researchers to find solutions for such an emerging threat. Future work will focus on studying the impact of systems like the one presented in this paper in technologies which use IPFS (e.g. blockchain) and the exploration of countermeasures (e.g. implementing an effective deletion system in IPFS).

\section*{Acknowledgements}
This work was supported by the European Commission under the Horizon 2020 Programme (H2020), as part of the project \emph{YAKSHA} (Grant Agreement no. 780498) and is based upon work from COST Action CA17124: \emph{DigForASP Digital forensics: evidence analysis via intelligent systems and practices} (European Cooperation in Science and Technology).

\end{document}